\documentclass[useAMS,usenatbib]{mn2e}
\usepackage{graphicx,natbib,times,amssymb}
\title[Heavily Obscured Quasars Host Galaxies are Disks]{Heavily Obscured Quasar Host Galaxies at $z\sim2$ are Disks, Not Major Mergers$^{1}$}
\author[Kevin Schawinski et al.]{
  \parbox[t]{16cm}{
Kevin Schawinski$^{2,3,6}$, 
Brooke D. Simmons$^{3,4}$,
C. Megan Urry$^{2,3,4}$,
Ezequiel Treister$^{5}$ and 
Eilat Glikman$^{3,4,7}$\\
  }\\
$^{1}$This work is based on observations taken by the CANDELS Multi-Cycle Treasury Program with the NASA/ESA HST, which is operated by the\\ Association of Universities for Research in Astronomy, Inc., under NASA contract NAS5-26555.\\
$^{2}$Department of Physics, Yale University, New Haven, CT 06520, U.S.A.\\
$^{3}$Yale Center for Astronomy and Astrophysics, Yale University, P.O. Box 208121, New Haven, CT 06520, U.S.A.\\
$^{4}$Department of Astronomy, Yale University, New Haven, CT 06520, U.S.A.\\
$^{5}$Universidad de Concepci\'{o}n, Departamento de Astronom\'{i}õa, Casilla 160-C, Concepci\'{o}n, Chile\\
$^{6}$Einstein Fellow\\
$^{7}$NSF Fellow
}

\begin{document}

\newcommand\aj{{AJ}}%
\newcommand\actaa{{Acta Astron.}}%
\newcommand\araa{{ARA\&A}}%
\newcommand\apj{{ApJ}}%
\newcommand\apjl{{ApJ}}%
\newcommand\apjs{{ApJS}}%
\newcommand\ao{{Appl.~Opt.}}%
\newcommand\apss{{Ap\&SS}}%
\newcommand\aap{{A\&A}}%
\newcommand\aapr{{A\&A~Rev.}}%
\newcommand\aaps{{A\&AS}}%
\newcommand\azh{{AZh}}%
\newcommand\baas{{BAAS}}%
\newcommand\caa{{Chinese Astron. Astrophys.}}%
\newcommand\cjaa{{Chinese J. Astron. Astrophys.}}%
\newcommand\icarus{{Icarus}}%
\newcommand\jcap{{J. Cosmology Astropart. Phys.}}%
\newcommand\jrasc{{JRASC}}%
\newcommand\memras{{MmRAS}}%
\newcommand\mnras{{MNRAS}}%
\newcommand\na{{New A}}%
\newcommand\nar{{New A Rev.}}%
\newcommand\pra{{Phys.~Rev.~A}}%
\newcommand\prb{{Phys.~Rev.~B}}%
\newcommand\prc{{Phys.~Rev.~C}}%
\newcommand\prd{{Phys.~Rev.~D}}%
\newcommand\pre{{Phys.~Rev.~E}}%
\newcommand\prl{{Phys.~Rev.~Lett.}}%
\newcommand\pasa{{PASA}}%
\newcommand\pasp{{PASP}}%
\newcommand\pasj{{PASJ}}%
\newcommand\qjras{{QJRAS}}%
\newcommand\rmxaa{{Rev. Mexicana Astron. Astrofis.}}%
\newcommand\skytel{{S\&T}}%
\newcommand\solphys{{Sol.~Phys.}}%
\newcommand\sovast{{Soviet~Ast.}}%
\newcommand\ssr{{Space~Sci.~Rev.}}%
\newcommand\zap{{ZAp}}%
\newcommand\nat{{Nature}}%
\newcommand\iaucirc{{IAU~Circ.}}%
\newcommand\aplett{{Astrophys.~Lett.}}%
\newcommand\apspr{{Astrophys.~Space~Phys.~Res.}}%
\newcommand\bain{{Bull.~Astron.~Inst.~Netherlands}}%
\newcommand\fcp{{Fund.~Cosmic~Phys.}}%
\newcommand\gca{{Geochim.~Cosmochim.~Acta}}%
\newcommand\grl{{Geophys.~Res.~Lett.}}%
\newcommand\jcp{{J.~Chem.~Phys.}}%
\newcommand\jgr{{J.~Geophys.~Res.}}%
\newcommand\jqsrt{{J.~Quant.~Spec.~Radiat.~Transf.}}%
\newcommand\memsai{{Mem.~Soc.~Astron.~Italiana}}%
\newcommand\nphysa{{Nucl.~Phys.~A}}%
\newcommand\physrep{{Phys.~Rep.}}%
\newcommand\physscr{{Phys.~Scr}}%
\newcommand\planss{{Planet.~Space~Sci.}}%
\newcommand\procspie{{Proc.~SPIE}}%
\newcommand\helvet{{Helvetica~Phys.~Acta}}%

\def\Chandra{\textit{Chandra}}
\def\XMM{\textit{XMM-Newton}}
\def\Swift{\textit{Swift}}

\def\OI{[\mbox{O\,{\sc i}}]~$\lambda 6300$}
\def\OIII{[\mbox{O\,{\sc iii}}]~$\lambda 5007$}
\def\SII{[\mbox{S\,{\sc ii}}]~$\lambda \lambda 6717,6731$}
\def\NII{[\mbox{N\,{\sc ii}}]~$\lambda 6584$}

\def\Ha{{H$\alpha$}}
\def\Hb{{H$\beta$}}

\def\NIIHa{[\mbox{N\,{\sc ii}}]/H$\alpha$}
\def\SIIHa{[\mbox{S\,{\sc ii}}]/H$\alpha$}
\def\OIHa{[\mbox{O\,{\sc i}}]/H$\alpha$}
\def\OIIIHb{[\mbox{O\,{\sc iii}}]/H$\beta$}

\def\Ebmv{E($B-V$)}
\def\LOIII{$L[\mbox{O\,{\sc iii}}]$}
\def\Ledd{${L/L_{\rm Edd}}$}
\def\LOIIIs4{$L[\mbox{O\,{\sc iii}}]$/$\sigma^4$}
\def\LOIIIMbh{$L[\mbox{O\,{\sc iii}}]$/$M_{\rm BH}$}
\def\Mbh{$M_{\rm BH}$}
\def\Msigma{$M_{\rm BH} - \sigma$}
\def\Ms{$M_{\rm *}$}
\def\Msun{$M_{\odot}$}
\def\Msunyr{$M_{\odot}yr^{-1}$}

\def\ergs{$~\rm erg~s^{-1}$}
\def\kms{$~\rm km~s^{-1}$}

\def\galfit{\texttt{GALFIT}}
\def\multidrizzle{\texttt{multidrizzle}}

\def\sersic{S\'{e}rsic}

\date{}

\pagerange{\pageref{firstpage}--\pageref{lastpage}} \pubyear{2012}

\maketitle

\label{firstpage}

\begin{abstract}
We explore the nature of heavily obscured quasar host galaxies at $z\sim2$ using deep \textit{Hubble Space Telescope} WFC3/IR imaging of 28 Dust Obscured Galaxies (DOGs) to investigate the role of major mergers in driving black hole growth. The high levels of obscuration of the quasars selected for this study act as a natural coronagraph, blocking the quasar light and allowing a clear view of the underlying host galaxy. The sample of heavily obscured quasars represents a significant fraction of the cosmic mass accretion on supermassive black holes as the quasars have inferred bolometric luminosities around the break of the quasar luminosity function. We find that only a small fraction (4\%, at most 11-25\%) of the quasar host galaxies are major mergers. Fits to their surface brightness profiles indicate that 90\% of the host galaxies are either disk dominated, or have a significant disk. This disk-like host morphology, and the corresponding weakness of bulges, is evidence against major mergers and suggests that secular processes are the predominant driver of massive black hole growth. Finally, we suggest that the co-incidence of mergers and AGN activity is luminosity dependent, with only the most luminous quasars being triggered mostly by major mergers.
\end{abstract}


\section{Introduction}
\label{sec:intro}
Our view on which processes are important in triggering and fuelling black hole accretion phases has have changed as the first rest-frame optical observations of $z\sim2$ active galactic nucleus (AGN) host galaxies are being made with the new WFC3/IR on the \textit{Hubble Space Telescope}. \cite{2011ApJ...727L..31S} used the Early Release Science observations of a sample of $1<z<3$ X-ray selected moderate-luminosity ( $10^{42} < L_{\rm X} < 10^{44}$\ergs) AGN to show that the majority of the host galaxies feature disk-dominated rest-frame optical light profiles, while showing no significant signs of mergers and interactions; this was recently confirmed by \cite{2012ApJ...744..148K} with a larger sample.  AGN selected in X-rays trace the peak of black hole growth at moderate luminosities and therefore capture a significant fraction of cosmic black hole growth that results in normal black holes at $z\sim0$, such as the black hole at the centre of the Milky Way \citep[e.g.][]{2003ApJ...598..886U, 2005A&A...441..417H}. The fact that a large fraction of this black hole growth is not associated with major mergers, but rather with disk host morphologies suggests that secular processes are most important in driving most black hole growth.

\begin{figure*}
\begin{center}

\includegraphics[width=0.99\textwidth]{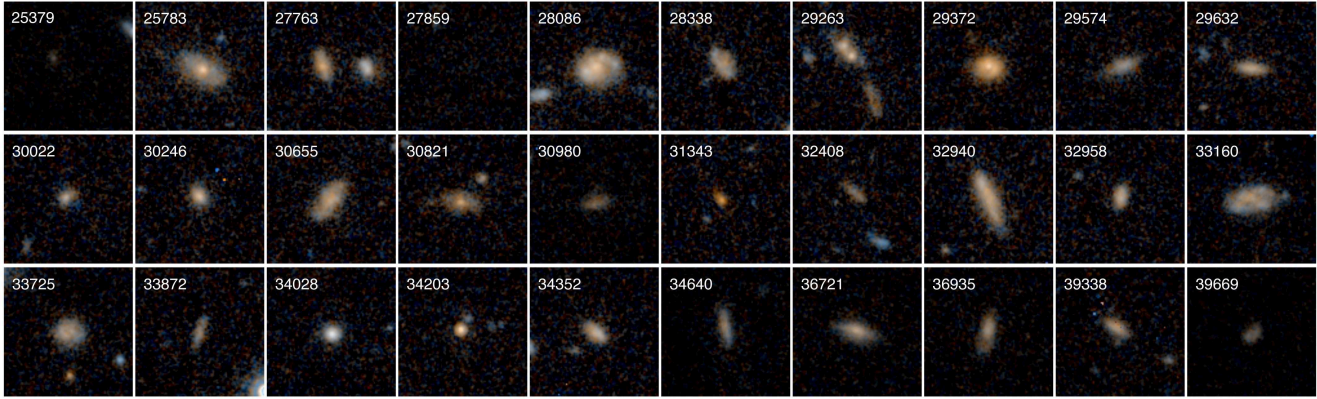}

\caption{Two-colour \textit{Hubble Space Telescope} WFC3/IR composite images of the sample of the 30 heavily obscured quasar host galaxies at $z\sim2$ presented in this \textit{Letter}. The composites consist of the deep CANDELS F125W (\textit{blue}) and F160W (\textit{red}) images and are created using an asinh stretch \citep[following ][]{2004PASP..116..133L}. Each image is labeled by the object's MUSYC ID number \citep{2010ApJS..189..270C}.  }

\label{fig:rogues_gallery}

\end{center}
\end{figure*}

Where does this leave major mergers as drivers of black hole growth and spheroid formation? The major merger picture was initially based on the most ultra-luminous starburst galaxies transitioning to powerful quasars. Specifically, simulations of gas-rich major mergers resulted in first obscured and then unobscured very luminous quasars. We now look at the host galaxies of heavily obscured quasar-luminosity AGN.

 The classic merger picture was first outlined by \cite{1988ApJ...325...74S}: two gas-rich galaxies undergo a major merger, which in turn triggers a powerful, infrared-luminous starburst as well as accretion onto the black hole. Once the accretion episode reaches quasar luminosity, the quasar by some means (radiation, outflows, winds, jets, etc.) drives out the gas, terminating both the starburst and the quasar. Theorists have used this sequence as the basis for detailed simulations of merger-driven quasars and quenching, and it has become a major component of our understanding of galaxy evolution \citep{2005Natur.435..629S, 2005Natur.433..604D, 2005ApJ...630..705H, 2006ApJS..163....1H, 2008ApJS..175..356H}.

Heavily obscured quasars are the best place to investigate the role of major mergers in triggering quasars, for two reasons: i)  the obscuration acts as a natural coronagraph, blocking the quasar light so we are allowed a clear view on the host galaxy, and ii)  simulations predict that the heavily obscured phase coincides with the peak of morphological disturbance as the progenitor galaxies are conflated in a powerful starburst. 

In this \textit{Letter}, we present new WFC3/IR rest-frame optical imaging of $z\sim2$ heavily obscured quasar host galaxies. The quasars selected have infrared and inferred bolometric luminosities suggesting that they are just below or at the `break' in the quasar luminosity function. This means that they represent a substantial fraction of cosmic black hole growth.

Throughout this \textit{Letter}, we assume a $\Lambda$CDM cosmology with $h_{0}=0.7$, $\Omega_{m}=0.27$ and $\Omega_{\Lambda}=0.73$, in agreement with the most recent cosmological observations \citep{2009ApJS..180..225H}.

\section{Observations}

\subsection{Sample Selection}

We select heavily obscured quasars in the Extended Chandra Deep Field South (E-CDFS) using the infrared excess method developed by \cite{2008ApJ...672...94F} \citep[see also ][]{2009ApJ...706..535T}. This method works by selecting intrinsically red objects using the observed  $R-K$ colour with excess mid-infrared emission seen in the $f(24\mu m)/f(R)$ ratio. For this study, we use the canonical selection criteria of $R-K >4.5$ and $f(24\mu m)/f(R) > 1000$. Objects selected this way are known as Dust Obscured Galaxies (DOGs).

From stacking of X-rays, we know that more than 90\% of DOGs must be Compton-thick ($N_{\rm H}$$>$10$^{24}$~cm$^{-2}$; $A_{\rm V}$$>$30-300) active galactic nuclei \citep{2008ApJ...672...94F, 2009ApJ...706..535T}. Since the sample used here is a random sub-sample of the one used by \cite{2009ApJ...706..535T}, we expect the AGN fraction to be comparable.

\subsection{\textit{Hubble Space Telescope} WFC3/IR Imaging Data}

The Chandra Deep Field South has been imaged over 39$\square\arcmin$ using the WFC3/IR F105W, F125W and F160W filters by the Cosmic Assembly Near-infrared Deep Extragalactic Legacy Survey (CANDELS;  \citealt{2011ApJS..197...36K, 2011ApJS..197...35G}). We obtained the 6-epoch mosaic (exposure time approx. 8 orbits) from the CANDELS website (\texttt{http://candels.ucolick.org/}) and cross-matched the F160W image with our DOG sample; this yields a total of 31 DOGs covered by the F160W image. We remove one object due to image artifacts. Because of the ongoing nature of the CANDELS observations, the exposure depth is not uniform across the field. We show postage stamps of the entire sample of 30 objects in Figure \ref{fig:rogues_gallery} and summarise their properties in Table \ref{tab:sample}.

\subsection{Mid-Infrared and Bolometric Luminosities}

We estimate the quasar bolometric luminosity in the DOG sample to place it in a cosmological context. Very few DOGs have spectroscopic redshifts \citep{2009ApJ...706..535T} but their photometric properties place them in the redshift range $1<z<3$. 

We convert the observed \textit{Spitzer} MIPS $24\mu m$ fluxes of the ECDFS DOGs from \cite{2009ApJ...706..535T} to intrinsic mid-IR luminosities by taking the median flux and assuming it arises from a source at $z=1$, 2 and 3, respectively, to span the plausible range. This yields observed $L_{24 \mu m} = 8.1 \times 10^{43}$, $4.6 \times 10^{44}$ and $1.2 \times 10^{45}$\ergs. Based on their observed (stacked) X-ray properties of DOGs, \cite{2009ApJ...706..535T} estimate typical bolometric luminosities of $L_{\rm bol} \sim 10^{45}$\ergs. Assuming a  \cite{1994ApJS...95....1E} quasar template redshifted to $z=1$, 2 and 3 yields bolometric luminosities of $L_{\rm bol} = 2.0 \times 10^{45}$, $1.1 \times 10^{46}$ and $7.4 \times 10^{46}$\ergs, respectively. These estimates of the bolometric luminosity are imprecise; the $24\mu m$ fluxes may contain at least some contribution from star formation, while the high levels of obscuration of the AGN indicates that even at $24\mu m$ (restframe $\sim 12\mu m$) some fraction of the intrinsic AGN luminosity remains suppressed.

The compilation of the quasar bolometric luminosity function by \cite*{2007ApJ...654..731H} places the DOGs in context: they report a break at $L^{*} =4.9 \times 10^{46}$\ergs\ at $z=2$ meaning the DOGs populate the break of the quasar luminosity function around the peak of accretion activity, or just below. The DOGs are thus part of the population where a substantial fraction of cosmic black hole growth occurs.

\section{Analysis}

We inspect the F160W images of the 30 obscured quasar host galaxies to determine whether they could plausibly be major mergers. The images are shown in Figure \ref{fig:rogues_gallery}. We perform two separate analyses of the images: visual inspection and parametric fits to the light profiles. Two objects are too faint in the F160W image to be visually analysed (25379 and 27859\footnote{Though we are able to fit 27859, with a resulting large error on the \sersic\ index.}).


\begin{figure}
\begin{center}

\includegraphics[width=0.45\textwidth]{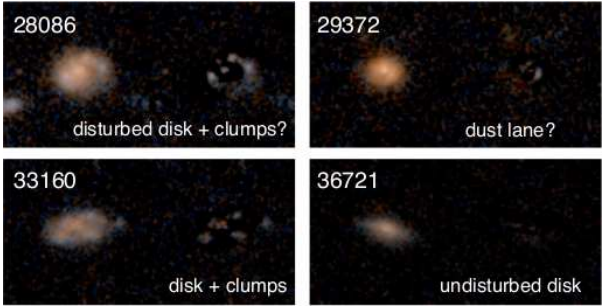}

\caption{Four example images (\textit{left}) and fit residuals (\textit{right}). The \textit{top-left} highlights the second most disturbed object (28086) whose residual image shows a minor disturbance. The \textit{top-right} is a galaxy (29372) whose original image could feature either two nuclei, or a dust lane. The residual image does not reveal two nuclei, so it is most likely a dust lane. 33160 in the \textit{bottom-left} is an example of a disk with (blue) star-forming clumps that become more prominent after the subtraction of a very disk-like \sersic\ model ($n_{\rm F160W}= 0.36\pm 0.03$). The \textit{bottom-right} is an example of a very clean subtraction of a disk-like \sersic\ model $n_{\rm F160W}=1.45\pm 0.06$, which does not reveal any hitherto invisible major merger signature such as double nuclei. }

\label{fig:residuals}

\end{center}
\end{figure}


\begin{figure}
\begin{center}

\includegraphics[angle=90,width=0.45\textwidth]{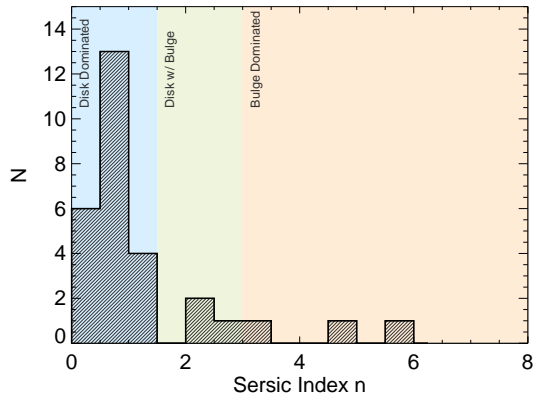}

\caption{The distribution of \sersic\ indices measured in the F160W ($H$-band) images of the heavily obscured quasar host galaxies (from Table \ref{tab:sample}). This histogram implies that 90\% of heavily obscured quasar host galaxies have significant disks and weak bulges. As shown in \citet{2008ApJ...683..644S}, the light of host galaxies with $n<1.5$ is dominated by a disk component. Intermediates at $1.5<n<3.0$ have a substantial disk, but also feature a bulge, while $n>3.0$ indicated bulge-dominated systems. }

 \label{fig:sersic}
 
\end{center}
\end{figure}

\begin{table*}
\begin{center}
\caption{The Sample of Dust-Obscured Galaxies (DOGs)}
\label{tab:sample}
\begin{tabular}{@{}lccclccl}
\hline
 \multicolumn{1}{l}{MUSYC} & \multicolumn{1}{c}{R.A.}  & \multicolumn{1}{c}{Dec.} & \multicolumn{1}{l}{$f(24\mu m)$}  & \multicolumn{1}{l}{Visual}  & \multicolumn{1}{c}{\sersic\ Index $n$} & \multicolumn{1}{c}{$L_{\rm{host}}/L_{\rm{PS}}$} & \multicolumn{1}{c}{Residual Image}\\
  \multicolumn{1}{l}{ID} & \multicolumn{1}{c}{(J2000)}  & \multicolumn{1}{c}{(J2000)} & \multicolumn{1}{l}{mJy}  & \multicolumn{1}{l}{Morphology$^{b}$} & \multicolumn{1}{c}{from F160W} & \multicolumn{1}{c}{Ratio$^{c}$} & \multicolumn{1}{c}{Appearance}\\
\hline
\hline
          25379    &  03:32:11.7 &-27:51:55.7    &      184.30  &  &     $ 2.74\pm 0.56$    &   $-$      &     \\
          25783    &  03:32:18.6 &-27:51:34.3    &      324.90  &  &     $ 1.14\pm 0.23$    &      0.91      &     \\
          27763    &  03:32:35.5 &-27:50:21.0    &       88.05  &  &     $ 1.00\pm 0.06$    &   $-$      &     \\
          27859    &  03:32:10.1 &-27:50:33.0    &      305.10  &  &             $-$    &   $-$      &     \\
          28086    &  03:32:17.5 &-27:50:03.0    &      321.90  &  merger?&     $ 0.81\pm 0.04$    &   $-$      & asymmetric residual$^{d}$    \\
          28338    &  03:32:44.8 &-27:49:54.0    &      155.00  &  &     $ 0.33\pm 0.06$    &   $-$      &     \\
          29263a    &  03:32:44.3 &-27:49:11.9    &      262.50  &  major merger&     $ 0.62\pm 0.23$    &      0.76      &     \\
          29263b    &    &  &       &  &     $ 1.69\pm0.53$    &       0.80      &     \\
          29372    &  03:32:35.7 &-27:49:15.9    &      569.30  & merger?  &     $ 0.51\pm 0.21$    &      0.89      &  asymmetric residual$^{d}$   \\
          29574    &  03:32:29.2 &-27:49:16.9    &      166.50  &  &     $ 0.90\pm 0.10$    &   $-$      &     \\
          29632    &  03:32:17.3 &-27:49:08.1    &       88.76  &  &     $ 0.76\pm 0.06$    &   $-$      &     \\
          30022    &  03:32:52.4 &-27:49:07.2    &       35.33  &  &     $ 2.49\pm 0.17$    &   $-$      &     \\
          30246    &  03:32:45.9 &-27:48:55.6    &       26.79  &  &     $ 2.06\pm 0.08$    &   $-$      &     \\
          30655    &  03:32:28.8 &-27:48:29.6    &      456.30  &  &     $ 0.78\pm 0.04$    &   $-$      &    faint SF clumps \\
          30821    &  03:32:24.3 &-27:48:30.6    &       92.29  &  faint neighbour&     $ 0.81\pm 0.24$    &      0.90      &     \\
          30980    &  03:32:05.8 &-27:48:20.0    &      219.80  &  &     $ 0.49\pm 0.03$    &   $-$      &     \\
          31343    &  03:32:42.9 &-27:48:09.4    &       32.93  &  &     $ 4.94\pm 0.82$    &   $-$      &     \\
          32408    &  03:32:50.1 &-27:47:33.0    &       71.51  &  &     $ 0.25\pm 0.04$    &   $-$      &     \\
          32940    &  03:32:34.4 &-27:46:59.6    &      120.10  &  &     $ 0.45\pm 0.04$    &   $-$      &     \\
          32958    &  03:32:49.6 &-27:47:14.9    &       99.77  &  &     $ 0.54\pm 0.04$    &   $-$      &     \\
          33160    &  03:32:04.9 &-27:46:47.3    &      613.60  &  &     $ 0.36\pm 0.03$    &   $-$      &   SF clumps  \\
          33725    &  03:32:06.8 &-27:46:43.6    &      247.30  &  &     $ 0.56\pm 0.05$    &   $-$      &     \\
          33872    &  03:32:21.1 &-27:46:44.0    &       26.93  &  &     $ 0.63\pm 0.06$    &   $-$      &     \\
         34028*    &  03:32:11.8 &-27:46:28.0    &      155.10  &  &     $ 3.18\pm 0.19$    &   $-$      &     \\
         34203*    &  03:32:38.0 &-27:46:26.4    &       93.26  & faint neighbour &     $ 5.66\pm 2.72$    &      0.42      &     \\
          34352    &  03:32:20.3 &-27:46:20.4    &       63.46  &  faint neighbour &     $ 0.74\pm 0.05$    &   $-$      &     \\
          34640    &  03:32:54.6 &-27:46:06.3    &       83.87  &  &     $ 0.81\pm 0.08$    &   $-$      &     \\
          36721    &  03:32:26.4 &-27:44:43.6    &      112.90  &  &     $ 1.45\pm 0.06$    &   $-$      &     \\
          36935    &  03:32:23.1 &-27:44:42.1    &       66.15  &  &     $ 1.12\pm 0.06$    &   $-$      &     \\
          39338    &  03:32:41.2 &-27:43:09.8    &       84.70  &  faint neighbour &     $ 0.83\pm 0.06$    &   $-$      &     \\
          39669    &  03:32:04.6 &-27:43:00.6    &      189.90  &  &     $ 0.23\pm 0.06$    &   $-$      &     \\
\hline
\end{tabular}
\\
\begin{flushleft}
$^a$ MUSYC catalog ID, see \cite{2010ApJS..189..270C}. Objects with X-ray detections are marked with $*$.\\
$^b$ See images shown in Figure \ref{fig:rogues_gallery}.\\
$^c$ The ratio of the host luminosity to the point source luminosity, reported only when \galfit\ requires an unresolved object to yield a physical fit. This may be due to an AGN point source (in the case of the X-ray-detected DOGs) or an unresolved bulge or central concentration, \textit{i.e.} a central bulge. \\
$^d$ See Figure \ref{fig:residuals}.
\end{flushleft}
\end{center}
\end{table*}


\subsection{Visual Inspection}

 From visual inspection, there is one clear major merger amongst the sample: 29263. It features two distinct components and a prominent tidal tail. It resembles simulations of gas-rich disk-disk mergers \citep[e.g.,][]{1996ApJ...471..115B, 2005Natur.433..604D}. Two further objects have features that can be interpreted as morphological disturbances: 28086 shows some asymmetry and 29372 may have two components, though this may be caused by a dust lane bisecting a single galaxy. We revisit both these objects in the following section and Figure \ref{fig:residuals}. 
 
 The remaining objects show no clear signs of major merger activity or bright nearby neighbours. Four objects (38021, 34203, 34352 and 39338) do show faint neighbours but we cannot determine whether they are associated in redshift. They do not appear to be connected by tidal tails or similar features. 

Removing the two objects that are too faint, this results in a major merger incidence of $1/28 = 4\%$ (assuming 29263 is the only true major merger) to $3/28 = 11\%$ (including the two potentially disturbed objects). For an extremely conservative limit on the merger fraction, we can include the four objects with faint neighbours, yielding a merger incidence of $7/28 = 25\%$. 

\subsection{\galfit\ Analysis}

We  analyse the F160W images of the DOGs using the \galfit\ surface brightness fitting tool \citep{2002AJ....124..266P}. This allows us to fit the \sersic\ indices of the obscured quasar host galaxies and to inspect the residuals after subtracting the model fits for signs of mergers, dust lanes or spiral arms that emerge only in the subtracted image. We choose the F160W band as the reddest image available to map the stellar distribution. Particular care is taken to fix the background level to that computed by \texttt{SExtractor} to avoid confusing the background with disk light profiles. One object (25379) was too faint to fit at all, while another (27859) is very faint, though we were able to fit it. The obvious major merger, 29263, was fit with separate \sersic\ models for the two nuclei and another for the tidal tail. The fits to 5 of the objects required the addition of a point source due to the presence of a central concentration. This could either be a bulge or a faint point source in the case of the two X-ray-detected DOGs. Note that the \textit{HST} WFC3/IR point spread function shape is similar to that of a small bulge, so we cannot tell the two apart at high redshift. However, the physical size of the PSF FWHM corresponds to $\sim1$ kpc, so in the case of an unresolved bulge, it is a minor component, and the whole galaxy is still disk-dominated. The results of this fitting analysis are shown in Table \ref{tab:sample} and examples shown in Figure \ref{fig:residuals}.

We find that the majority of obscured quasar host galaxies have low \sersic\ indices indicating a significant disk component to the restframe optical light profile. Following the simulations of  \cite{2008ApJ...683..644S}, the measured \sersic\ indices mean that 23 are pure disks, 3 have significant disks, and 3 are bulge-dominated. We show the distribution of measured \sersic\ indices in Figure \ref{fig:sersic}. Most of the residual images show very little leftover light after subtraction of the \sersic\ model. None of them reveal previously undetected double nuclei. In the case of host galaxies with a clumpy appearance, notably 28086 and 33160, the more prominent, star-forming clumps remain visible in the residual; they feature characteristic blue colours. We conclude that the surface brightness fits agree with the visual inspection in finding a very low incidence of major mergers and no otherwise hidden mergers. 

Finally, we verify that we are not missing bulges due to dust obscuration of bulge light \citep{2010MNRAS.403.2053G} by fitting a sample of star-forming galaxies with similar $K$-band magnitudes and redshifts as the DOGs and restframe $U-V<1.3$ and $V-J < 1.1$ in order to select relatively dust-free galaxies and find that 14/15 are disk dominated $n<1.5$ and one has some bulge ($n=2.61$).


\begin{figure}
\begin{center}

\includegraphics[angle=90,width=0.22\textwidth]{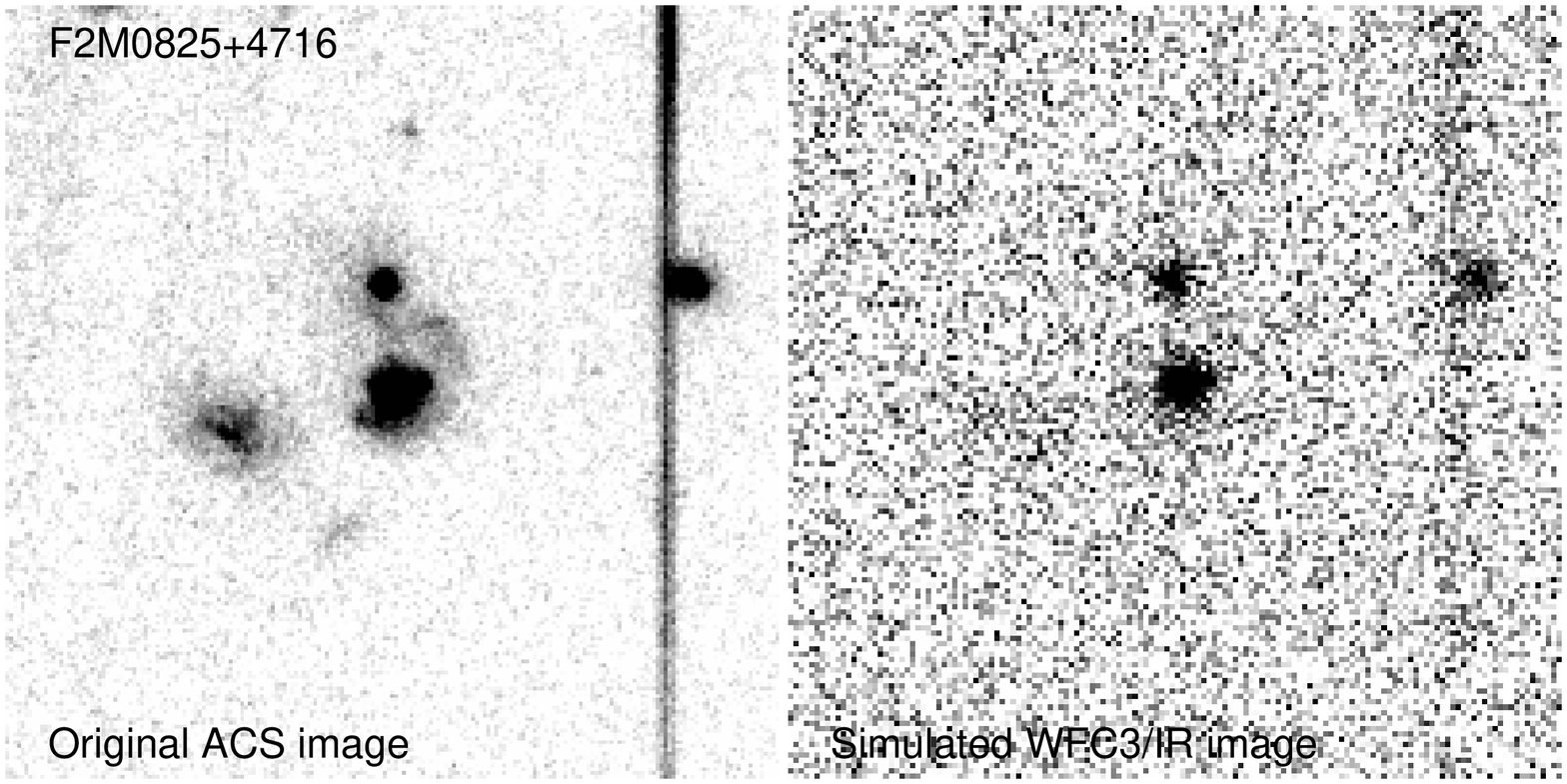}
\includegraphics[angle=90,width=0.22\textwidth]{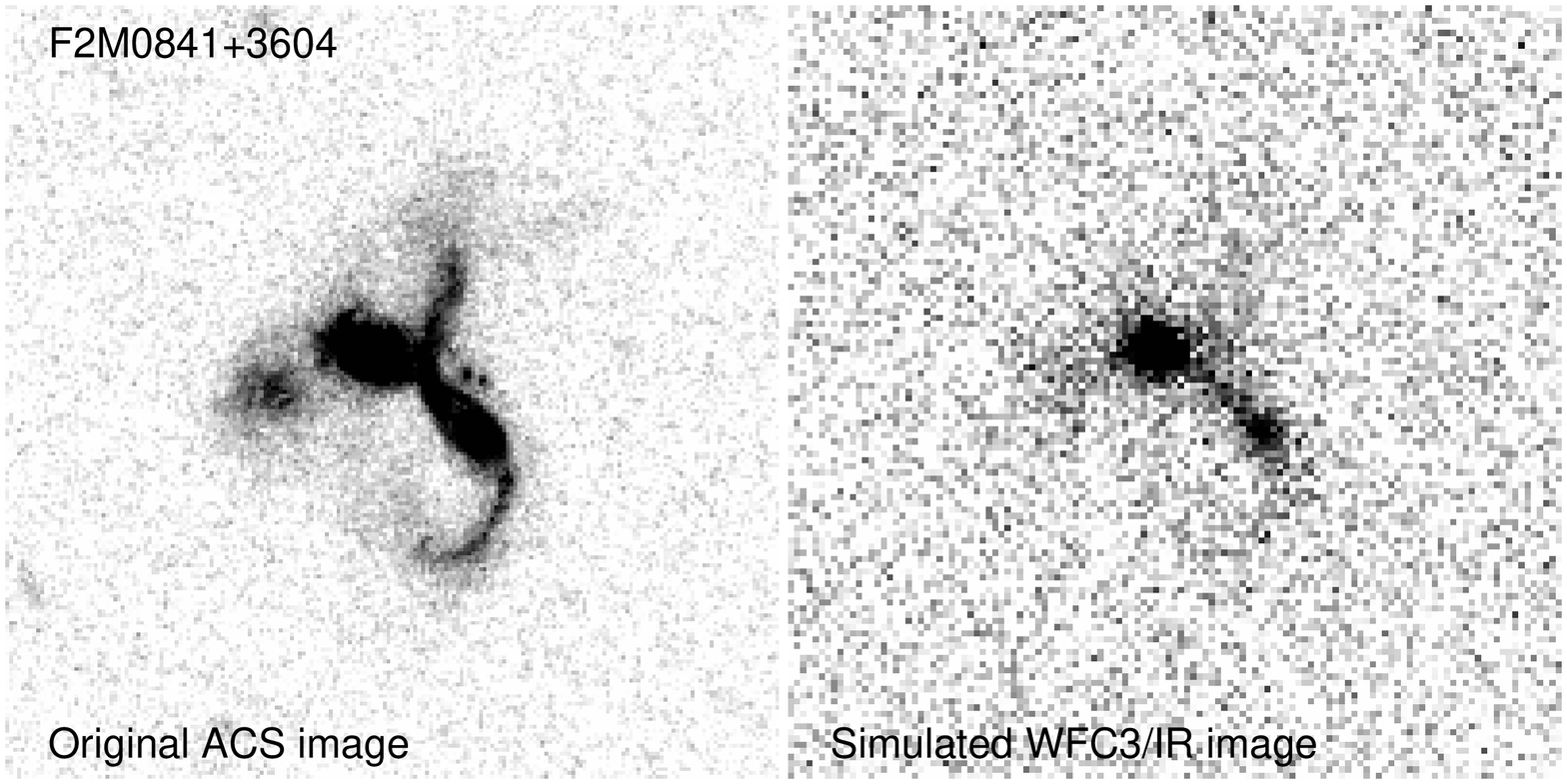}\\
\vspace{0.04in}
\includegraphics[angle=90,width=0.22\textwidth]{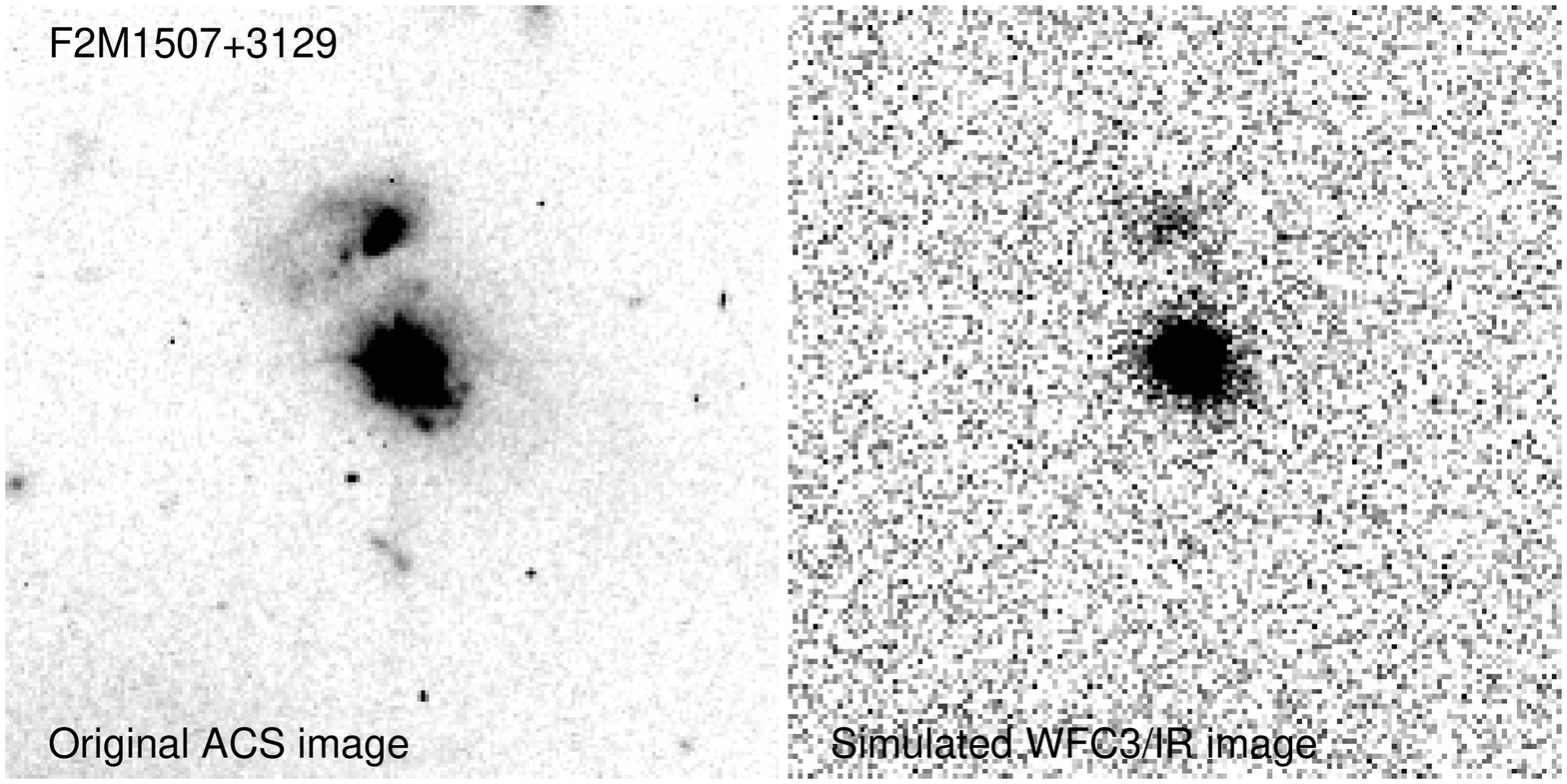}
\includegraphics[angle=90,width=0.22\textwidth]{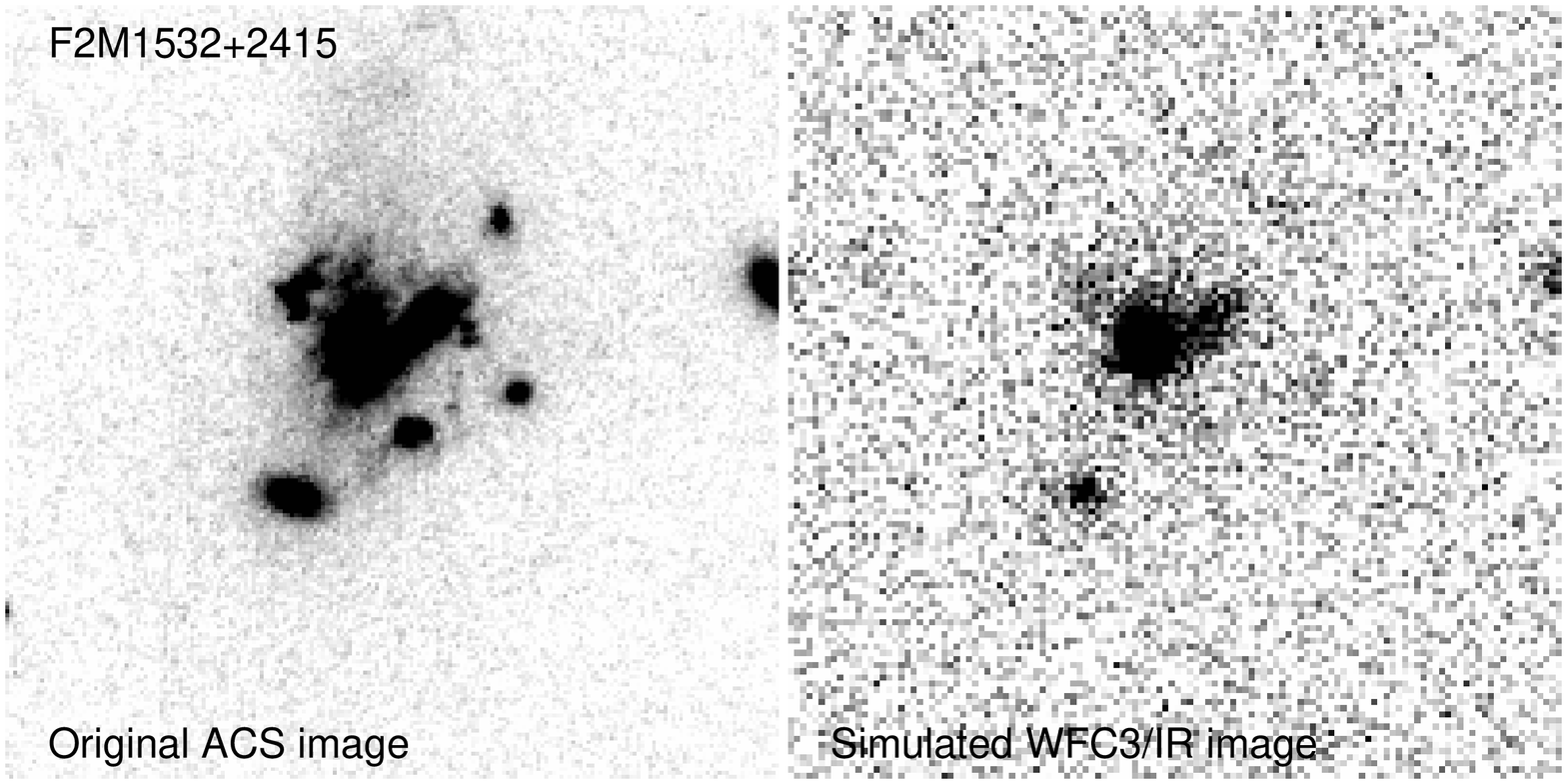}

\caption{Four examples of images of quasars in major mergers that were degraded to mimic the \textit{HST} WFC3/IR data in the CANDELS survey. The purpose of these simulated images is to determine whether features indicating an ongoing major merger would be visible in DOGs. The original images are \textit{HST} ACS F814W ($I$-band) images of $0.5<z<1.0$ red quasars obtained by \citet{2008ApJ...674...80U}, modified  using the following procedure: i) we use the $I$-band which corresponds to the $H$-band at $z\sim2$ (redshifted), ii) we match the pixel scale to that of the (drizzled) CANDELS data and then iii) convolve it with an empirical point spread function constructed from the CANDELS image; iv) we then add an appropriate level of noise. In each panel, we show the original ACS image with the quasar name (\textit{left}) and the simulated image (\textit{right}). This exercise shows that, while tidal tails and delicate disturbed features can disappear, the irregular, multi-component nature of major mergers remains visible.}

\label{fig:degraded}

\end{center}
\end{figure}

\subsection{Could We Miss Major Mergers?}

The main caveat with this analysis is the question of whether features indicative of a major merger would be apparent given the quality of the data. We approach this question empirically by taking \textit{Hubble Space Telescope} Advanced Camera for Surveys (ACS) $I$-band images of red quasars hosted by major mergers at $z<1$ \citep{2008ApJ...674...80U} and redshift them to $z\sim2$, where the observed $I$-band corresponds to the $H$-band. We change the pixel scale and convolve the images with the WFC3/IR point spread function (PSF) of the CANDELS observations. We also increase the background to result in comparable signal to noise. From this test, shown in Figure \ref{fig:degraded}, we find that the redshifted images lack some of the fine structure and faint tidal features but the major clumps and components that make up the major merger remain clearly visible as separate sources. This exercise shows that, while tidal tails and delicate disturbed features can disappear, the irregular, multi-component nature of major mergers remains visible. We conclude that, were the DOGs major mergers like the red quasars, we would have been able to detect this.

\subsection{Is the Merger Fraction Luminosity-Dependent?}

The heavily obscured quasars studied here do not represent the most luminous quasars in the Universe. The most luminous quasars at all redshifts are red quasars, type 1 quasars reddened by dust (\citealt{2012AAS...21920903G}; Glikman et al., submitted). Observations of red quasars at moderate redshift ( $0.5<z<1$) by  \cite{2008ApJ...674...80U} show a very high fraction, close to 100\%, of major mergers. Future observations of even more luminous red quasars at $z\sim2$ may well reveal similarly high levels of morphological disturbance. Observations of lower-luminosity active galaxies, on the other hand, show very low levels of major merger activity \citep[e.g.][]{2011ApJ...741L..11C, 2011ApJ...727L..31S,2012ApJ...744..148K}.  Given these observations, together with new new analysis presented here, we suggest that the role of major mergers in triggering black hole growth is a function of the bolometric luminosity, with major mergers being the main channel only at the highest luminosities.

\section{Discussion}

We have analysed deep \textit{Hubble Space Telescope} WFC3/IR imaging of a sample of 28 DOGs at $1<z<3$ , $\sim$90\% of which are expected to harbour a heavily obscured quasar whose infrared and inferred bolometric luminosities are just below or around the break in the quasar luminosity function. The rest-frame optical host galaxy morphologies indicate that only one is unambiguously a major merger. Three additional objects may be undergoing some disturbance, but their appearance can be accounted for by star-forming clumps or dust lanes and another four objects have faint neighbours, which may or may not be associated. We further analyse the F160W images using \galfit\ and verify that the residuals to the host galaxy fits do not reveal any hidden features or disturbances. Indeed, the host galaxies are smooth with low \sersic\ indices, which means that the host galaxies are either disk-dominated system or have substantial disks.

We assume that the merger fraction of the heavily obscured quasars in our sample is representative of their less obscured and unobscured counterparts. As simulations predict that the heavily obscured phase coincides with the early, most disturbed `train wreck' phase of the merger \citep[e.g.][]{2006ApJS..163....1H}, it may in fact be a high estimate.

These observational results challenge the picture in which quasar activity is triggered by major mergers. We find that the heavily obscured quasars studied here cannot be in the early- to mid stages of a merger, as double nuclei and perhaps tidal tails would be apparent, as they are in one case. They also cannot represent the final stages of a merger, as 90\% have low \sersic\ indices due to disk-dominated light profiles; the major merger should have disrupted the disk and built a spheroid by this stage. Simulations show that the merger remnants can re-grow a disk after some time \citep[e.g.][]{2006ApJ...645..986R}, though it requires extremely gas-rich mergers with particular initial orbits.

In summary, black hole growth in heavily obscured quasars near the break in the luminosity function at $z\sim2$, the peak epoch of black hole growth, occurs predominantly in disk galaxies rather than in major  `trainwreck' mergers. As the break in the luminosity function is where most black hole growth occurs, our result  implies that secular processes, rather than major mergers, are the predominant driver of cosmic massive black hole growth.


\section*{Acknowledgements}
Support for the work of KS was provided by NASA through Einstein Postdoctoral Fellowship grant numbers PF9-00069, issued by the Chandra X-ray Observatory Center, which is operated by the SAO for and on behalf of NASA under contract NAS8-03060.BDS acknowledges support Yale University and from NASA  grant HST-AR-12638.01-A. ET received partial support from CATA-BASAL (PFB 06) and FONDECYT grant 1120061. This research has made use of NASA's ADS Service.


\bibliographystyle{mn}

\bsp

\label{lastpage}

\end{document}